\documentclass[twocolumn,showpacs,preprintnumbers,amsmath,amssymb,pre]{revtex4}

\usepackage{graphicx}
\usepackage{dcolumn}
\usepackage{bm}
\usepackage{enumerate}

\begin{document}

\title{Subdiffusion in a system with a partially permeable partially absorbing wall}

\author{Tadeusz Koszto{\l}owicz}
 \email{tadeusz.kosztolowicz@ujk.edu.pl}
 \affiliation{Institute of Physics, Jan Kochanowski University,\\
         ul. Uniwersytecka 7, 25-406 Kielce, Poland}

\date{\today}

\begin{abstract}
We consider subdiffusion of a particle in a one-dimensional system with a thin partially permeable wall. Passing through the wall, the particle can be absorbed with a certain probability. We call such a wall partially permeable partially absorbing wall (PPAW). Using the diffusion model in a system with discrete time and spatial variable, probability densities (Green's functions) describing subdiffusion in the system have been derived. Knowing the Green's functions we derive boundary conditions at the wall. The boundary conditions take a specific form in which time derivatives of the fractional order controlled by the subdiffusion parameter are involved. We assume that the absorption of a particle can occur only when the particle jumps through the wall. It is not possible to temporarily retain a particle inside a thin wall. The wall can represent a thin membrane. If a system with a thick membrane inside which particles may diffuse is considered, it can be treated as a three-part with a thick membrane as the middle part. The boundary conditions at membrane surfaces can be assumed as for PPAW. The system with PPAW can be used to filter diffusing particles. The temporal evolution of the probability that the diffusing molecule has not been absorbed is considered. This function shows the efficiency of the filtering process. The impact of subdiffusion and wall parameters on this function is discussed.
\end{abstract}

\maketitle

\section{Introduction\label{sec1}}

In many processes in physics, biology, engineering sciences and others normal diffusion or subdiffusion occur in a system with a thin membrane which is partially permeable and partially absorbing for diffusing particles \cite{hsieh,hobbie,luckey}. Absorption is defined here as the permanent exclusion of a particle from further diffusion. It can be, for example, the disintegration of the particle or its permanent immobilization. Some membranes can be used as filters in the process of recovering substances. The examples are diffusion dialysis in which acids are recovered from aqueous solutions \cite{lan,luo,lin1,wang,vasileva,ye}, carbon dioxide recovery \cite{lin,cui}, absorption of substances in a rubber membrane \cite{mccarley}, recover of hydrochloride acid from waste pickling solution \cite{gueccia}, the use of membranes as filters for sterilization and purification of protein pharmaceuticals \cite{brose} and elimination of ethanol \cite{kalant}, diffusion of drugs through membranes made of regenerated keratin and ceramides \cite{selmin}, diffusion of various substances through a partially absorbing skin \cite{hansen}. Various scenarios for retaining diffusing particles through a filter membrane have been considered, for example sieve filtration and molecular filtration \cite{brose,lee}, and diffusion of particles with the multi-stage surface reactions, such as viral entry into healthy cells \cite{chou}. Diffusing particles can be trapped, among others, at the membrane surface \cite{prager}, by absorbing patches on membrane \cite{lawley}, and by binding sites on the wall of a membrane channel \cite{db2006}.  

When modelling diffusion in a system with a partially absorbing thin membrane, one needs two boundary conditions at the membrane to solve diffusion equation. These boundary conditions are usually assumed or derived using a model of particle transport through a membrane. The boundary conditions at the membrane may contain integrodifferential operator with a memory kernel. Many boundary conditions at a thin membrane not equivalent to each other have been considered, see for example the discussion in \cite{tkarxiv2020} and the references cited therein. Theoretical models often use diffusion-reaction equation or diffusion equation with reactions on the membrane surface at which radiative boundary condition have been often supposed \cite{pal,rn,vko,goychuk}. Radiative boundary condition assumes that the particles flux at the absorbing surface is proportional to the particles concentration at the surface. 

We consider subdiffusion in a one--dimensional system with a partially permeable partially absorbing wall (PPAW). PPAW is defined here as a thin obstacle for diffusing particles inside which the particle cannot stop. A particle that tries to jump over PPAW can do it with some probability or be stopped on the wall surface. The particle can be absorbed with some probability as it passes through the wall. The wall represents thin membrane which can absorb diffusing particles. If a system with a thick membrane inside which particles may diffuse is considered, the system can be treated as a three-part with a thick membrane as the middle part. The membrane surfaces can be treated as PPAWs. 

Subdiffusion occurs in media in which particle jumps are strongly hindered due to a complex structure of the medium \cite{ks,mk,mk1}. The example are subdiffusion in gels \cite{kdm,bm}, biological cells \cite{bf,bpf,hf}, biological membranes \cite{mjc,jjm,coker}, and in media having a fractal structure \cite{iom,slz}. Within the Continuous Time Random Walk model waiting time for the particle to jump is anomalously long for subdiffusion; the probability density distribution of this time $\psi$ has a heavy tail, $\psi(t)\sim 1/t^{\alpha+1}$, $t\rightarrow\infty$, $0<\alpha<1$, which leads to infinite mean value of this time \cite{ks,mk}. Subdiffusion is usually described by a subdiffusion equation with fractional time derivative of the order controlled by subdiffusion parameter $\alpha$, see Eq. (\ref{eq1}) in Sec. \ref{sec2}.

Models of a particle random walk in a discrete system are useful to derive boundary conditions \cite{vko,goychuk,lzm,zlm,kb,kb1,tk1,tk2,tk3,tk4}. In some models, it is assumed that inside the membrane a particle must stop for some time. Absorption may occur during the stay of the molecule inside the membrane. The probability of absorption then depends on the absorption coefficient as well as the distribution of time that the molecule stay inside the membrane \cite{vko,goychuk,lzm,zlm,kb,kb1}. We have different situation in the model considered in this paper. A particle may, with some probability, jump over the wall or be stopped on its surface. It is not possible to retain the particle inside the wall. Similar model of subdiffusion with non--absorbing thin membrane has been already considered \cite{tk1,tk2,tk3,tk4}. However, the presence of a partially absorbing membrane in the system changes the dynamics of the subdiffusion process and provides qualitatively different Green's functions and boundary conditions at the wall. We derive the Green's functions and boundary conditions according to the rule: {\it a particle can be absorbed only when it jumps through the wall}. The probability that the molecule will not be absorbed in the time interval $(0,t)$ depends on the number of particle jumps through the membrane as well as on the probabilities of absorption during a particle jump. In deriving the Green's function, we take into account the number of jumps through the wall. As far as we know, this issue has not been considered yet. We mention that  processes depending on the number of visits of the diffusing particle at a specific point in the system have been considered \cite{montroll65,weiss}. In this paper we consider the process depending on the number of jumps between two different points separated by a partially permeable wall.

\section{Method\label{sec2}}

We assume that subdiffusion parameters do not depend on time and spatial variable. Let $P(x,t;x_0)$ be a probability density (the Green's function) of finding a diffusing particle at the point $x$ at time $t$, $x_0$ is the initial location of the particle. The probability fulfils the subdiffusion equation 
\begin{eqnarray}\label{eq1}
	\frac{\partial P(x,t;x_0)}{\partial t}=D\frac{\partial^{1-\alpha}}{\partial t^{1-\alpha}}\frac{\partial^2 P(x,t;x_0)}{\partial x^2}\;,
\end{eqnarray}
with the initial condition $P(x,0;x_0)=\delta(x-x_0)$, where $\delta$ is the Dirac delta function, $D$ is the subdiffusion coefficient, $\alpha$ is the subdiffusion parameter, the Riemann--Liouville fractional derivative occurring in Eq. (\ref{eq1}) is defined for $\beta>0$ as
\begin{equation}\label{eq2}
\frac{d^\beta f(t)}{dt^\beta}=\frac{1}{\Gamma(n-\beta)}\frac{d^n}{dt^n}\int_0^t dt'(t-t')^{n-\beta-1}f(t'),
\end{equation}
where $n=[\beta]+1$, $[\beta]$ is the integral part of $\beta$.

In further considerations we use the Laplace transform $\mathcal{L}[f(t)]=\int_0^\infty{\rm e}^{-st}f(t)dt\equiv\hat{f}(s)$. Using the formula 
\begin{equation}\label{eq3}
\mathcal{L}\left[\frac{d^{\beta}f(t)}{d t^{\beta}}\right]=s^\beta \hat{f}(s),
\end{equation}
$0<\beta<1$, we get the subdiffusion equation in terms of the Laplace transform
\begin{eqnarray}\label{eq4}
	s\hat{P}(x,s;x_0)-P(x,0;x_0)=Ds^{1-\alpha}\frac{\partial^2 \hat{P}(x,s;x_0)}{\partial x^2}\;.
\end{eqnarray}

We consider subdiffusion in a system with a thin partially permeable partially absorbing wall (PPAW) which can absorb a diffusing particle when it is passing through the wall. In the following considerations the key is to find the probability densities $P^{(k_A,k_B)}(x,t;x_0)$ that a particle performs $k_A$ passages from the left side of the wall to its right side and $k_B$ passages when it moves in opposite direction. The numbers $k_A$ and $k_B$ may differ by at most 1, $|k_A-k_B|\leq 1$. The probabilities fulfil the relation 
\begin{equation}\label{eq5}
P(x,t;x_0)=\sum_{k_A,k_B=0}^\infty P^{(k_A,k_B)}(x,t;x_0).
\end{equation} 
Let $\rho(k_A,k_B)$ be the probability that the particle has not been absorbed after passing $k_A$ and $k_B$ times through the wall. The probability density $P(x,t;x_0|\rho)$ that the particle is at position $x$ at time $t$ and has not been absorbed up to time $t$ is
\begin{equation}\label{eq6}
P(x,t;x_0|\rho)=\sum_{k_A,k_B=0}^\infty \rho(k_A,k_B) P^{(k_A,k_B)}(x,t;x_0).
\end{equation} 

We assume that an absorption probability of particle which passes through the wall located at $x_N$ from the region $A=(-\infty,x_N)$ to the region $B=(x_N,\infty)$ is $1-\rho_A$ and an absorption probability for particle moving in opposite direction is $1-\rho_B$, $\rho_A$ and $\rho_B$ are the probabilities that the particle is not absorbed during its passage through the wall. We suppose that these probabilities do not change over time and do not depend on the number of particle passing through the wall. The probability that the particle still exists in the system after making $k_A$ and $k_B$ passing through the wall reads
\begin{equation}\label{eq7}
\rho(k_A,k_B)=\rho_A^{k_A}\rho_B^{k_B}.
\end{equation}
In the following we label functions describing subdiffusion by the lower indexes $i,j\in\{A,B\}$ that denote in which region a point is located, the first index $i$ denotes the location of $x$ and the second one $j$ the location of $x_0$. In the following, we assume that $x_0$ is in the region $A$. To shorten the notation we use the parameter $k$, $k\equiv k_A$, which means the number of particle passages through the wall from region $A$ to $B$. Then, the number of particle passes from region $B$ to $A$ is $k_B=k$ if $x\in A$ or $k_B=k-1$ if $x\in B$.
We have
\begin{eqnarray}\label{eq8}
P_{AA}(x,t;x_0|\rho)=\sum_{k=0}^\infty \rho^k_A\rho^k_B P_{AA}^{(k,k)}(x,t;x_0)\;,
\end{eqnarray}
\begin{eqnarray}\label{eq9}
P_{BA}(x,t;x_0|\rho)=\sum_{k=1}^\infty \rho^k_A\rho_B^{k-1} P_{BA}^{(k,k-1)}(x,t;x_0)\;.
\end{eqnarray}

To derive the functions $P_{AA}^{(k,k)}$ and $P_{BA}^{(k,k-1)}$ it is convenient to use a particle random walk model in a system with non--absorbing wall and with discrete time and spatial variable.

\section{Particle random walk in a discrete system with a non--absorbing partially permeable wall\label{sec3}} 

\begin{figure}[htb]
\centerline{%
\includegraphics[width=7cm]{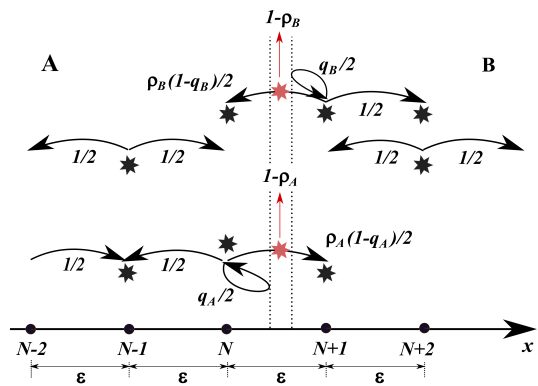}}
\caption{The system with a partially permeable partially absorbing wall located between $N$ and $N+1$ sites. $1-\rho_A$ and $1-\rho_B$ are the probabilities that the particle will be absorbed during its jump through the wall. A particle that tries to get from region $A$ to $B$ through the wall can do it with probability $\rho_A(1-q_A)/2$, and moving in the opposite direction with probability $\rho_B(1-q_B)/2$. The probabilities that a particle which tries to pass the wall will be stopped are $q_A$ and $q_B$, these probabilities are independent of the absorption probabilities.}
\label{Fig1}
\end{figure}

The model we use is based on difference equations and differs from the classical Continuous Time Random Walk (CTRW) model, because the random variable is only the time between consecutive jumps of the particle. In the CTRW model both a length of particle jump and a time which is needed to take a particle step are random variables. The partially permeable wall is located between $N$ and $N+1$ sites. The distance between discrete sites $\varepsilon$ is treated as a small parameter. The regions are defined here as $A=(-\infty,N]$ and $B=[N+1,\infty)$. A particle that tries to jump through the wall from $N$ to $N+1$ site can do it with the probability $(1-q_A)/2$ or can be stopped by the wall with the probability $q_A/2$. We make similar assumptions when the particle tries to jump from $N+1$ to $N$, see Fig. \ref{Fig1}. 

We denote as $P_{ij,n}(m;m_0)$, $i,j\in\{A,B\}$, a probability that the particle is located at $m$ in region $i$ after $n$ steps, $m_0$ in region $j$ is the initial position of the particle. We assume that $m_0$ is located in the region $A$, the initial conditions read $P_{AA,0}(m;m_0)=\delta_{m,m_0}$ and $P_{BA,0}(m;m_0)=0$. 
The difference equations describing the particle random walk in a discrete system with a thin partially permeable wall are
\begin{eqnarray}
\label{eq10}P_{AA,n+1}(m;m_0)=\frac{1}{2}P_{AA,n}(m-1;m_0)\\
 +\frac{1}{2}P_{AA,n}(m+1;m_0),\;m<N,\nonumber\\ 
      \nonumber\\
\label{eq11}P_{AA,n+1}(N;m_0)=\frac{1}{2}P_{AA,n}(N-1;m_0)\\ 
+\frac{q_A}{2}P_{AA,n}(N;m_0)+\frac{1-q_B}{2}P_{BA,n}(N+1;m_0),\nonumber\\ 
      \nonumber\\
\label{eq12}P_{BA,n+1}(N+1;m_0)=\frac{1-q_A}{2}P_{AA,n}(N;m_0)\\
+\frac{q_B}{2}P_{BA,n}(N+1;m_0)\nonumber
+\frac{1}{2}P_{BA,n}(N+2;m_0),\nonumber
			\nonumber\\
\label{eq13}P_{BA,n+1}(m;m_0)=\frac{1}{2}P_{BA,n}(m-1;m_0)\\
 +\frac{1}{2}P_{BA,n}(m+1;m_0),\;m>N+1,\nonumber 
\end{eqnarray}

We are going to find the probabilities $P_{AA,n}^{(k,k)}(m;m_0)$ and $P_{BA,n}^{(k,k-1)}(m,m_0)$. Let $F_{ij,n}(m,m_0)$ be the probability that particle starting form $m_0$ in the region $j$ reaches the point $m$ in the region $i$ first time at the $n$-th step, $i,j\in\{A,B\}$. When $m,m_0\leq N$, the movement of the particle can be decomposed into the following stages for $k\geq 1$:
\begin{enumerate}
	\item\label{i1} the particle starts from $m_0$ and reaches the point $N$ first time after $n_{0A}$ steps with probability $F_{AA,n_{0A}}(N,m_0)$,
	\item\label{i2} starting from $N$ the particle passes through the PPAW and reaches the point $N+1$ first time at $n_{1A}$--th step with probability
\begin{equation}\label{eq14}
V_{A,n_{1A}}\equiv F_{BA,n_{1A}}(N+1,N),
\end{equation}
returning back the particle reaches the point $N$ first time after leaving it at $n_{1B}$-th step with probability
\begin{equation}\label{eq15} 
V_{B,n_{1B}}\equiv F_{AB,n_{1B}}(N,N+1),
\end{equation}
the situation described in this point is repeated $k$ times with the number of steps between successive passes through the PPAW denoted as $n_{1A}, n_{1B}, n_{2A}, n_{2B}, \ldots, n_{kA}, n_{kB}$,
	\item\label{i3} after the last pass through the PPAW the particle come form $N$ to $m$ without passing through the wall with probability $P^{(0,0)}_{AA,n-n_{0A}-n_{1A}-n_{1B}\ldots-n_{kA}-n_{kB}}(m;N)$.
\end{enumerate}
The function $P_{AA,n}^{(k,k)}$, $k\geq 1$, is the convolution of the functions described in the points \ref{i1}--\ref{i3}, the convolution is defined as $f_n*g_n\equiv \sum_{j=0}^n f_{n-j}g_j$. When $m_0\leq N$ and $m\geq N+1$ we only take into account the points \ref{i2} and \ref{i3}, but in the point \ref{i2} if the passage through the membrane from the point $N$ to $N+1$ occurs $k$ times, there are $k-1$ passes in the opposite direction and in the point \ref{i3} we change the function $P_{AA,n}^{(0,0)}$ to $P_{BB,n}^{(0,0)}$. We get
\begin{eqnarray}\label{eq16}
P^{(k,k)}_{AA,n}(m;m_0)=F_{AA,n}(N;m_0)\\ 
*\underbrace{(V_{A,n}*V_{B,n})*\ldots *(V_{A,n}*V_{B,n})}_{\rm {k\;times}}
 \;*P^{(0,0)}_{AA,n}(m;N),\nonumber
\end{eqnarray}
\begin{eqnarray}\label{eq17}
P^{(k,k-1)}_{BA,n}(m;m_0)=F_{AA,n}(N;m_0)\\
*\underbrace{(V_{A,n}*V_{B,n})*\ldots *(V_{A,n}*V_{B,n})}_{\rm {k-1\;times}}\;*V_{A,n} \nonumber\\
 *P^{(0,0)}_{BB,n}(m;N+1),\nonumber
\end{eqnarray}
where $k\geq 1$. 

Using the generating function 
\begin{equation}\label{eq18}
S_P(m,z;m_0)= \sum_{n=0}^\infty z^n P_n(m;m_0), 
\end{equation}
we move to continuous time by means of the equation \cite{ks,mk,montroll65}
\begin{equation}\label{eq19}
\hat{P}(m,s;m_0)=S_P(m,\hat{\psi}(s);m_0)\frac{1-\hat{\psi}(s)}{s},
\end{equation}
where $\hat{\psi}(s)$ is the Laplace transform of the distribution of time which is needed to take particle next step. 
Since $S_{f*g}=S_f \cdot S_g$, we get from Eqs. (\ref{eq16})--(\ref{eq18})
\begin{eqnarray}\label{eq20}
S_{P_{AA}^{(k,k)}}(m,z;m_0)=S_{F_{AA}}(N,z;m_0)\\ 
\times V^k_A(z)V^k_B(z) S_{P_{AA}^{(0,0)}}(m,z;N),\nonumber
\end{eqnarray}
\begin{eqnarray}\label{eq21}
S_{P_{BA}^{(k,k-1)}}(m,z;m_0)=S_{F_{AA}}(N,z;m_0)\\ 
\times V^k_A(z)V^{k-1}_B(z) S_{P_{BB}^{(0,0)}}(m,z;N+1),\nonumber
\end{eqnarray}
where $V_i(z)=\sum_{n=1}^\infty z^n V_{i,n}$, $i=A,B$.

The generating function $S_{F_{ij}}(m;m_0)=\sum_{n=1}^\infty z^n F_{ij,n}(m;m_0)$ can be calculated by means of the formula \cite{comment}
\begin{equation}\label{eq22}
S_{F_{ij}}(m,z;m_0)=\frac{S_{P_{ij}}(m,z;m_0)-\delta_{m,m_0}}{S_{P_{ij}}(m,z;m)},
\end{equation}
where $S_{P_{ij}}$ denotes the generating functions for the solutions to Eqs. (\ref{eq10})--(\ref{eq13}); these generating functions are presented in the Appendix Eqs. (\ref{a1})--(\ref{a4}).
We get
\begin{eqnarray}
V_A(z)\equiv S_{F_{BA}}(N+1,z;N)=\frac{(1-q_A)\eta(z)}{1-q_A\eta(z)},\label{eq23}\\
V_B(z)\equiv S_{F_{AB}}(N,z;N+1)=\frac{(1-q_B)\eta(z)}{1-q_B\eta(z)},\label{eq24}
\end{eqnarray}
and
\begin{eqnarray}\label{eq25}
 S_{F_{AA}}(N,z;m_0)= \left\{\begin{array}{ll}
      \eta^{N-m_0}(z)\;,m_0<N,\\
			   \\
      \frac{(1+\eta(z))(1-q_B\eta(z))}{\sqrt{1-z^2}\left[1-(q_A+q_B-1)\eta(z)\right]},m_0=N,
    \end{array}\right.
\end{eqnarray}
where
\begin{equation}\label{eq26}
\eta(z)=\frac{1-\sqrt{1-z^2}}{z}.
\end{equation}
The form of Eq. (\ref{eq25}) suggests that the cases of $m_0<N$ and $m_0=N$ should be considered separately. However, we note that the latter case is included in the function $V_A$ which describes the particle first passage time from $N$ to $N+1$ site. In this case the function $S_{F_{AA}}(N,z;N)$ should be omitted in Eq. (\ref{eq20}). In the following we consider the case of $m_0<N$. The obtained results will be also valid for the case of $m_0=N$.

The method of deriving the generating functions that the particle moves from $m_0$ to $m$ without passing through the membrane is described in Appendix. The functions are
\begin{eqnarray}\label{eq27}
S_{P^{(0,0)}_{AA}}(m,z;m_0)=\frac{[\eta(z)]^{|m-m_0|}}{\sqrt{1-z^2}}\\
+\left(\frac{q_A-\eta(z)}{1-q_A\eta(z)}\right)\frac{[\eta(z)]^{2N-m-m_0+1}}{\sqrt{1-z^2}},\nonumber
\end{eqnarray}
\begin{eqnarray}\label{eq28}
S_{P^{(0,0)}_{BB}}(m,z;m_0)=\frac{[\eta(z)]^{|m_0-m|}}{\sqrt{1-z^2}}\\
+\left(\frac{q_B-\eta(z)}{1-q_B\eta(z)}\right)\frac{[\eta(z)]^{m+m_0-2N-1}}{\sqrt{1-z^2}}.\nonumber
\end{eqnarray}

To move to continuous time we use Eq. (\ref{eq19}). After calculations we get
\begin{eqnarray}\label{eq29}
P_{AA}^{(0,0)}(m,s;m_0)=\frac{1-\hat{\psi}(s)}{s\sqrt{1-\hat{\psi}^2(s)}}\Bigg(\left[\eta(\hat{\psi}(s))\right]^{|m-m_0|}\\
+\left[\eta(\hat{\psi}(s))\right]^{2N-m-m_0+1}\frac{q_A-\eta(\hat{\psi}(s))}{1-q_A\eta(\hat{\psi}(s))}\Bigg),\nonumber
\end{eqnarray}
\begin{eqnarray}\label{eq30}
P_{AA}^{(k,k)}(m,s;m_0)=\frac{1-\hat{\psi}(s)}{s\sqrt{1-\hat{\psi}^2(s)}}\left[\eta(\hat{\psi}(s))\right]^{2N-m-m_0}\\
\times V^k_A(\hat{\psi}(s)) V^k_B(\hat{\psi}(s))\frac{1-\left[\eta(\hat{\psi}(s))\right]^2}{1-q_A\eta(\hat{\psi}(s))},\nonumber
\end{eqnarray}
\begin{eqnarray}\label{eq31}
P_{BA}^{(k,k-1)}(m,s;m_0)=\frac{1-\hat{\psi}(s)}{s\sqrt{1-\hat{\psi}^2(s)}}V^k_A(\hat{\psi}(s))\\ 
\times V_B^{k-1}(\hat{\psi}(s))\frac{1-\left[\eta(\hat{\psi}(s))\right]^2}{1-q_B\eta(\hat{\psi}(s))}\left[\eta(\hat{\psi}(s))\right]^{m-m_0-1},\nonumber
\end{eqnarray}
$k\geq 1$. Moving from a discrete position $m$ to a continuous spatial variable $x$ we use the equations $x=\varepsilon m$, $x_0=\varepsilon m_0$, $x_N=\varepsilon m_N$, and $P(x,t;x_0)=P(m,t;m_0)/\varepsilon$, where $\varepsilon$ is the distance between discrete sites. Next, we take the limit of small $\varepsilon$. The function $\hat{\psi}$ is assumed to be
\begin{equation}\label{eq32}
\hat{\psi}(s)=\frac{1}{1+\varepsilon^2\frac{s^\alpha}{2D}}.
\end{equation}
The motivation of Eq. (\ref{eq32}) is as follows. It is shown in \cite{tk1} that both the subdiffusion equation Eq. (\ref{eq1}) and the Green's function for homogeneous system can be derived from the discrete model only if $\hat{\psi}$ is expressed by Eq. (\ref{eq32}). Taking into account the first--order terms with respect to $\varepsilon$, we have 
\begin{eqnarray}\label{eq33}
\eta(\hat{\psi}(s))=1-\varepsilon\sqrt{\frac{s^\alpha}{D}}\;.
\end{eqnarray}
For $q_A$ and $q_B$ independent of $\epsilon$ we obtain $V_i(\hat{\psi}(s))\rightarrow 1$ when $\varepsilon\rightarrow 0$. In this case the first passage times from $N$ to $N+1$ and from $N+1$ to $N$ are independent of permeability properties of the wall. The reason is that, due to the formula $\nu(t)=\mathcal{L}^{-1}[\hat{\psi}(s)/(1-\hat{\psi}(s))]=2Dt^{1-\alpha}/\Gamma(\alpha)\varepsilon^2$ where $\nu$ is the frequency of jumps between adjacent sites, $\nu(t)\rightarrow\infty$ when $\varepsilon\rightarrow 0$. Thus, in any time interval a particle makes infinite number of attempts to pass the wall. If the wall is partially permeable, $0<q_A,q_B<1$, the probability that particle which tries to pass the wall does it is equal to one. Then, the wall loses its selective property. To avoid this non--physical result we assume that the permeability coefficients $1-q_A$ and $1-q_B$ depend on the parameter $\varepsilon$. Guided by the similar discussion in \cite{tk1,tk4}, we assume that $1-q_i=\varepsilon^{\mu_i}\sigma_i$, $i=A,B$. Then, we obtain
\begin{equation}\label{eq34}
V_i(\hat{\psi}(s))=\frac{\sigma_i-\varepsilon\sigma_i\sqrt{\frac{s^\alpha}{D}}}{\sigma_i-\varepsilon\sigma_i\sqrt{\frac{s^\alpha}{D}}+\varepsilon^{\mu_i-1}\sqrt{\frac{s^\alpha}{D}}}\;.
\end{equation}
We get finite $V_i$ which depends on $\sigma_i$ in the limit $\varepsilon\rightarrow 0$ only if $\mu_i=1$. Thus, we have
\begin{equation}\label{eq35}
\sigma_i=\frac{1-q_i}{\varepsilon},
\end{equation}
and for $\varepsilon\rightarrow 0$ we obtain
\begin{eqnarray}\label{eq36}
\hat{V}_i(\hat{\psi}(s))=\frac{\sigma_i}{\sigma_i+\sqrt{\frac{s^\alpha}{D}}},
\end{eqnarray}
$i=A,B$. From the above equations we get
\begin{eqnarray}\label{eq37}
\hat{P}_{AA}^{(0,0)}(x,s;x_0)=\frac{s^{\alpha/2-1}}{2\sqrt{D}}\Bigg[{\rm e}^{-\sqrt{\frac{s^\alpha}{D}}|x-x_0|}\\
+\frac{\sqrt{\frac{s^\alpha}{D}}-\sigma_A}{\sqrt{\frac{s^\alpha}{D}}+\sigma_A}\;{\rm e}^{-\sqrt{\frac{s^\alpha}{D}}(2x_N-x-x_0)}\Bigg],\nonumber
\end{eqnarray}
\begin{eqnarray}\label{eq38}
\hat{P}_{AA}^{(k,k)}(x,s;x_0)=\frac{s^{\alpha-1}}{\sigma_A D}\hat{V}^k_A(\hat{\psi}(s))\hat{V}^k_B(\hat{\psi}(s))\\
 \times \;{\rm e}^{-\sqrt{\frac{s^\alpha}{D}}(2x_N-x-x_0)},\;k\geq 1,\nonumber
\end{eqnarray}
\begin{eqnarray}\label{eq39}
\hat{P}_{BA}^{(k,k)}(x,s;x_0)=\frac{s^{\alpha-1}}{\sigma_B D}\hat{V}^k_A(\hat{\psi}(s))\hat{V}^{k-1}_B(\hat{\psi}(s))\\
 \times \;{\rm e}^{-\sqrt{\frac{s^\alpha}{D}}(x-x_0)},\;k\geq 1,\nonumber
\end{eqnarray}
with $V_A$ and $V_B$ expressed by Eq. (\ref{eq36}).

\section{Subdiffusion in a system with partially permeable partially absorbing wall\label{sec4}}

To find the Green's functions for a system with a partially permeable partially absorbing wall, we use Eqs. (\ref{eq8}) and (\ref{eq9}). Knowing Green's functions we derive boundary conditions at the wall. Finally, we find the temporal evolution of the probability that a diffusing particle still exists in the system. This function shows how effective the filtration process is. 
 
\subsection{Green's functions\label{sec4a}}

From Eqs. (\ref{eq8}), (\ref{eq9}), and (\ref{eq37})--(\ref{eq39}) we get
\begin{eqnarray}\label{eq40}
\hat{P}_{AA}(x,s;x_0|\rho)=\frac{\sqrt{s^\alpha}}{2s\sqrt{D}}\Big[\;{\rm e}^{-\sqrt{\frac{s^\alpha}{D}}|x-x_0|}\\
+\Big(\Xi_{A0}(s)+\Xi_{AA}(s|\rho)\Big){\rm e}^{-\sqrt{\frac{s^\alpha}{D}}(2x_N-x-x_0)}\Big],\nonumber
\end{eqnarray}
\begin{eqnarray}\label{eq41}
\hat{P}_{BA}(x,s;x_0|\rho)=\frac{\sqrt{s^\alpha}}{2s\sqrt{D}}\;\Xi_{BA}(s|\rho)\\
\times{\rm e}^{-\sqrt{\frac{s^\alpha}{D}}(x-x_0)},\nonumber
\end{eqnarray}
where 
\begin{equation}\label{eq42}
\Xi_{A0}(s)=\frac{\sqrt{\frac{s^\alpha}{D}}-\sigma_A}{\sqrt{\frac{s^\alpha}{D}}+\sigma_A},
\end{equation}
\begin{eqnarray}\label{eq43}
\Xi_{AA}(s|\rho)=\frac{2}{\sigma_A}\sqrt{\frac{s^\alpha}{D}}\sum_{k=1}^\infty \rho^k_A\rho^k_B\\
\times V_A^{k+1}(\hat{\omega}(s))V_B^k(\hat{\omega}(s)) =\frac{2}{\sigma_A\left(1+\frac{1}{\sigma_A}\sqrt{\frac{s^\alpha}{D}}\right)}\sqrt{\frac{s^\alpha}{D}}\nonumber\\ 
\times\frac{\rho_A\rho_B}{\left(1+\frac{1}{\sigma_A}\sqrt{\frac{s^\alpha}{D}}\right)\left(1+\frac{1}{\sigma_B}\sqrt{\frac{s^\alpha}{D}}\right)-\rho_A\rho_B},\nonumber
\end{eqnarray}
\begin{eqnarray}\label{eq44}
\Xi_{BA}(s|\rho)=\frac{2}{\sigma_B}\sqrt{\frac{s^\alpha}{D}}\sum_{k=1}^\infty \rho^k_A\rho^{k-1}_B\\ 
\times V_A^k(\hat{\omega}(s))V_B^k(\hat{\omega}(s))=\frac{2}{\sigma_B}\sqrt{\frac{s^\alpha}{D}}\nonumber\\ 
\times\frac{\rho_A}{\left(1+\frac{1}{\sigma_A}\sqrt{\frac{s^\alpha}{D}}\right)\left(1+\frac{1}{\sigma_B}\sqrt{\frac{s^\alpha}{D}}\right)-\rho_A\rho_B}.\nonumber
\end{eqnarray}

In order to calculate the inverse Laplace transform of the obtained functions we present $\Xi_{A0}$, $\Xi_{AA}$, and $\Xi_{BA}$ in the form of power series with respect to $s$, and then we use the equation
\begin{eqnarray}\label{eq45}
\mathcal{L}^{-1}\left[s^\nu {\rm e}^{-as^\beta}\right]\equiv f_{\nu,\beta}(t;a)\\
=\frac{1}{t^{\nu+1}}\sum_{k=0}^\infty{\frac{1}{k!\Gamma(-k\beta-\nu)}\left(-\frac{a}{t^\beta}\right)^k}\;,\nonumber
\end{eqnarray}
$a,\beta>0$; the function $f_{\nu,\beta}$ can be expressed by the Wright function and the H-Fox function, see Appendix.
Due to the formula $1/[1+au(1+bu)]=\sum_{n=0}^\infty d_n u^n$, $d_n=\sum_{i=0}^{[n/2]} {n-i\choose i} (-b/a)^i$, we obtain
\begin{eqnarray}\label{eq46}
\Xi_{A0}(s)=-1-2\sum_{n=1}^\infty \left(-\frac{1}{\sigma_A}\sqrt{\frac{s^\alpha}{D}}\right)^n,
\end{eqnarray}
\begin{equation}\label{eq47}
\Xi_{AA}(s|\rho)=\frac{2\rho_A\rho_B}{\sigma_A(1-\rho_A\rho_B)}\sum_{n=0}^\infty c_n\left(\sqrt{\frac{s^\alpha}{D}}\right)^{n+1},
\end{equation}
\begin{equation}\label{eq48}
\Xi_{BA}(s|\rho)=\frac{2\rho_A}{\sigma_B(1-\rho_A\rho_B)}\sum_{n=0}^\infty d_n\left(\sqrt{\frac{s^\alpha}{D}}\right)^{n+1},
\end{equation}
where 
\begin{eqnarray}
\label{eq49}c_n&=&\sum_{j=0}^n(-1/\sigma_A)^{n-j}d_j,\\
\label{eq50}d_n&=&\sum_{i=0}^{[n/2]} {n-i\choose i}\tilde{\sigma}^i(-\xi)^{n-i},  
\end{eqnarray}
$\tilde{\sigma}=1/(\sigma_A+\sigma_B)$, and $\xi=(1/\sigma_A+1/\sigma_B)/(1-\rho_A\rho_B)$.

Combining Eqs. (\ref{eq40}), (\ref{eq41}), and (\ref{eq46})--(\ref{eq50}) we get the Laplace transform of Green's functions
\begin{eqnarray}\label{eq51}
\hat{P}_{AA}(x,s;x_0|\rho)=\frac{\sqrt{s^\alpha}}{2s\sqrt{D}}\Big[{\rm e}^{-\sqrt{\frac{s^\alpha}{D}}|x-x_0|}\\
-\;{\rm e}^{-\sqrt{\frac{s^\alpha}{D}}(2x_N-x-x_0)}\Big]\nonumber\\
+\frac{1}{s}\sum_{n=0}^\infty g_n s^{\alpha(n+2)/2}\;{\rm e}^{-\sqrt{\frac{s^\alpha}{D}}(2x_N-x-x_0)},\nonumber
\end{eqnarray}
\begin{eqnarray}\label{eq52}
\hat{P}_{BA}(x,s;x_0|\rho)=\frac{1}{s}\sum_{n=0}^\infty h_n s^{\alpha(n+2)/2}\\
\times\;{\rm e}^{-\sqrt{\frac{s^\alpha}{D}}(x-x_0)},\nonumber
\end{eqnarray}
where
\begin{eqnarray}
\label{eq53}g_n&=&\frac{1}{\sigma_A(\sqrt{D})^{n+2}}\left[\left(-\frac{1}{\sigma_A}\right)^n+\frac{\rho_A\rho_B}{1-\rho_A\rho_B}c_n\right],\\
\label{eq54}h_n&=&\frac{\rho_A d_n}{(1-\rho_A\rho_B)\sigma_B(\sqrt{D})^{n+2}}.
\end{eqnarray}
Calculating the inverse Laplace transform of Eqs. (\ref{eq51}) and (\ref{eq52}) using Eq. (\ref{eq45}) term by term, we obtain
\begin{eqnarray}\label{eq55}
P_{AA}(x,t;x_0|\rho)=\frac{1}{2\sqrt{D}}\Bigg[f_{\alpha/2-1,\alpha/2}\left(t;\frac{|x-x_0|}{\sqrt{D}}\right)\\
-f_{\alpha/2-1,\alpha/2}\left(t;\frac{2x_N-x-x_0}{\sqrt{D}}\right)\nonumber\\
+\sum_{n=0}^\infty g_n f_{(n+2)\alpha/2-1,\alpha/2}\left(t;\frac{2x_N-x-x_0}{\sqrt{D}}\right)\Bigg]\nonumber,
\end{eqnarray}
\begin{eqnarray}\label{eq56}
P_{BA}(x,t;x_0|\rho)=\sum_{n=0}^\infty h_n f_{(n+2)\alpha/2-1,\alpha/2}\left(t;\frac{x-x_0}{\sqrt{D}}\right).
\end{eqnarray}

The Green's functions for the system with non--absorbing membrane $P(x,t;x_0|1)$ can be obtained putting $\rho_A=\rho_B=1$ in the above equations. After calculations we get
\begin{eqnarray}\label{eq57}
\hat{P}_{AA}(x,s;x_0|1)=\frac{\sqrt{s^\alpha}}{2s\sqrt{D}}\Big[{\rm e}^{-\sqrt{\frac{s^\alpha}{D}}|x-x_0|}\\
+{\rm e}^{-\sqrt{\frac{s^\alpha}{D}}(2x_N-x-x_0)}\Big]\nonumber\\
-\frac{\sigma_A}{s}\sum_{n=0}^\infty \left(-\frac{\sqrt{s^\alpha}}{\sqrt{D}(\sigma_A+\sigma_B)}\right)^{n+1}{\rm e}^{-\sqrt{\frac{s^\alpha}{D}}(2x_N-x-x_0)},\nonumber
\end{eqnarray}
\begin{eqnarray}\label{eq58}
\hat{P}_{BA}(x,s;x_0|1)=\frac{\sigma_A}{s}\sum_{n=0}(-1)^n \left(\frac{\sqrt{s^\alpha}}{\sqrt{D}(\sigma_A+\sigma_B)}\right)^{n+1}\\
\times\;{\rm e}^{-\sqrt{\frac{s^\alpha}{D}}(x-x_0)}.\nonumber
\end{eqnarray}
In the time domain we have
\begin{eqnarray}\label{eq59}
P_{AA}(x,t;x_0|1)=\frac{1}{2\sqrt{D}}\Bigg[f_{\alpha/2-1,\alpha/2}\left(t;\frac{|x-x_0|}{\sqrt{D}}\right)\\
+f_{\alpha/2-1,\alpha/2}\left(t;\frac{2x_N-x-x_0}{\sqrt{D}}\right)\Bigg]\nonumber\\
+\sum_{n=0}^\infty \left(-\frac{1}{\sqrt{D}(\sigma_A+\sigma_B)}\right)^{n+1}\nonumber\\
\times f_{(n+1)\alpha/2-1,\alpha/2}\left(t;\frac{2x_N-x-x_0}{\sqrt{D}}\right)\nonumber,
\end{eqnarray}
\begin{eqnarray}\label{eq60}
P_{BA}(x,t;x_0|1)=\sigma_A\sum_{n=0}^\infty \left(-\frac{1}{\sqrt{D}(\sigma_A+\sigma_B)}\right)^{n+1}\\
\times f_{(n+1)\alpha/2-1,\alpha/2}\left(t;\frac{x-x_0}{\sqrt{D}}\right).\nonumber
\end{eqnarray}

\begin{figure}[htb]
\centerline{%
\includegraphics[width=10cm]{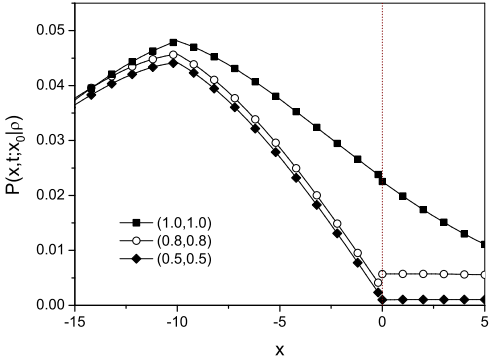}}
\caption{The plots of Green's functions for different parameters $(\rho_A,\rho_B)$ given in the legend. The other parameters are $\alpha=0.9$, $D=10$, $t=5$, $\sigma_A=\sigma_B=2$, $x_0=-10$, and $x_N=0$. The parameters are given in arbitrarily chosen units.}
\label{Fig2}
\end{figure}

Examples of plots of the Green's functions Eqs. (\ref{eq55}), (\ref{eq56}), (\ref{eq59}), and (\ref{eq60}) are presented in Fig. \ref{Fig2}.

\subsection{Boundary conditions at a partially permeable partially absorbing wall\label{sec4b}}

\begin{figure}[htb]
\centerline{%
\includegraphics[width=10cm]{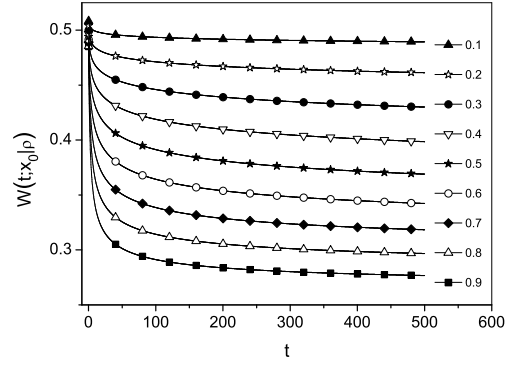}}
\caption{The plots of the function $W$ Eq. (\ref{eq72}) for different $\alpha$ given in the legend, $\rho_A=\rho_B=0.5$, $\sigma_A=\sigma_B=2$, and $D=5$.}
\label{Fig3}
\end{figure}
\begin{figure}[htb]
\centerline{%
\includegraphics[width=10cm]{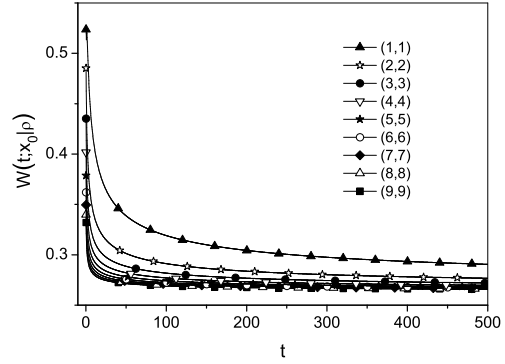}}
\caption{The plots of $W$ for different $(\sigma_A,\sigma_B)$ given in the legend, $\alpha=0.9$, $\rho_A=\rho_B=0.5$, and $D=5$.}
\label{Fig4}
\end{figure}
\begin{figure}[htb]
\centerline{%
\includegraphics[width=10cm]{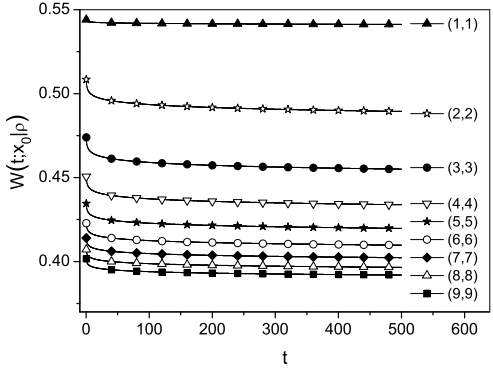}}
\caption{The plots of $W$ for different $(\sigma_A,\sigma_B)$ given in the legend, $\alpha=0.1$, the other parameters are the same as in Fig. \ref{Fig4}.}
\label{Fig5}
\end{figure}
\begin{figure}[htb]
\centerline{%
\includegraphics[width=10cm]{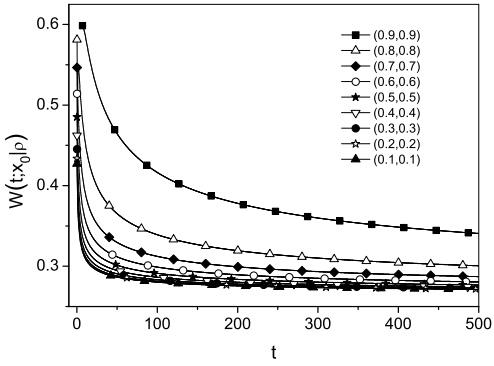}}
\caption{The plots of $W$ for different $(\rho_A,\rho_B)$ given in the legend, $\alpha=0.9$, $\sigma_A=\sigma_B=2$, and $D=5$.}
\label{Fig6}
\end{figure}
\begin{figure}[htb]
\centerline{%
\includegraphics[width=10cm]{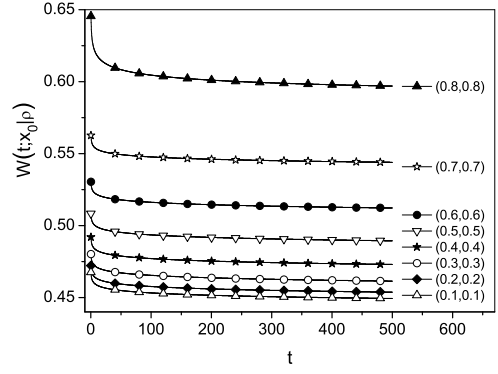}}
\caption{The plots of $W$ for different $(\rho_A,\rho_B)$ given in the legend, $\alpha=0.1$, the other parameters are the same as in Fig. \ref{Fig6}.}
\label{Fig7}
\end{figure}
\begin{figure}[htb]
\centerline{%
\includegraphics[width=10cm]{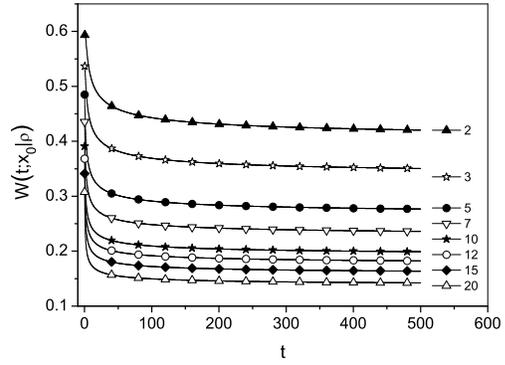}}
\caption{The plots of $W$ for different $D$ given in the legend, $\alpha=0.9$, $\rho_A=\rho_B=0.5$, and $\sigma_A=\sigma_B=2$.}
\label{Fig8}
\end{figure}
\begin{figure}[htb]
\centerline{%
\includegraphics[width=10cm]{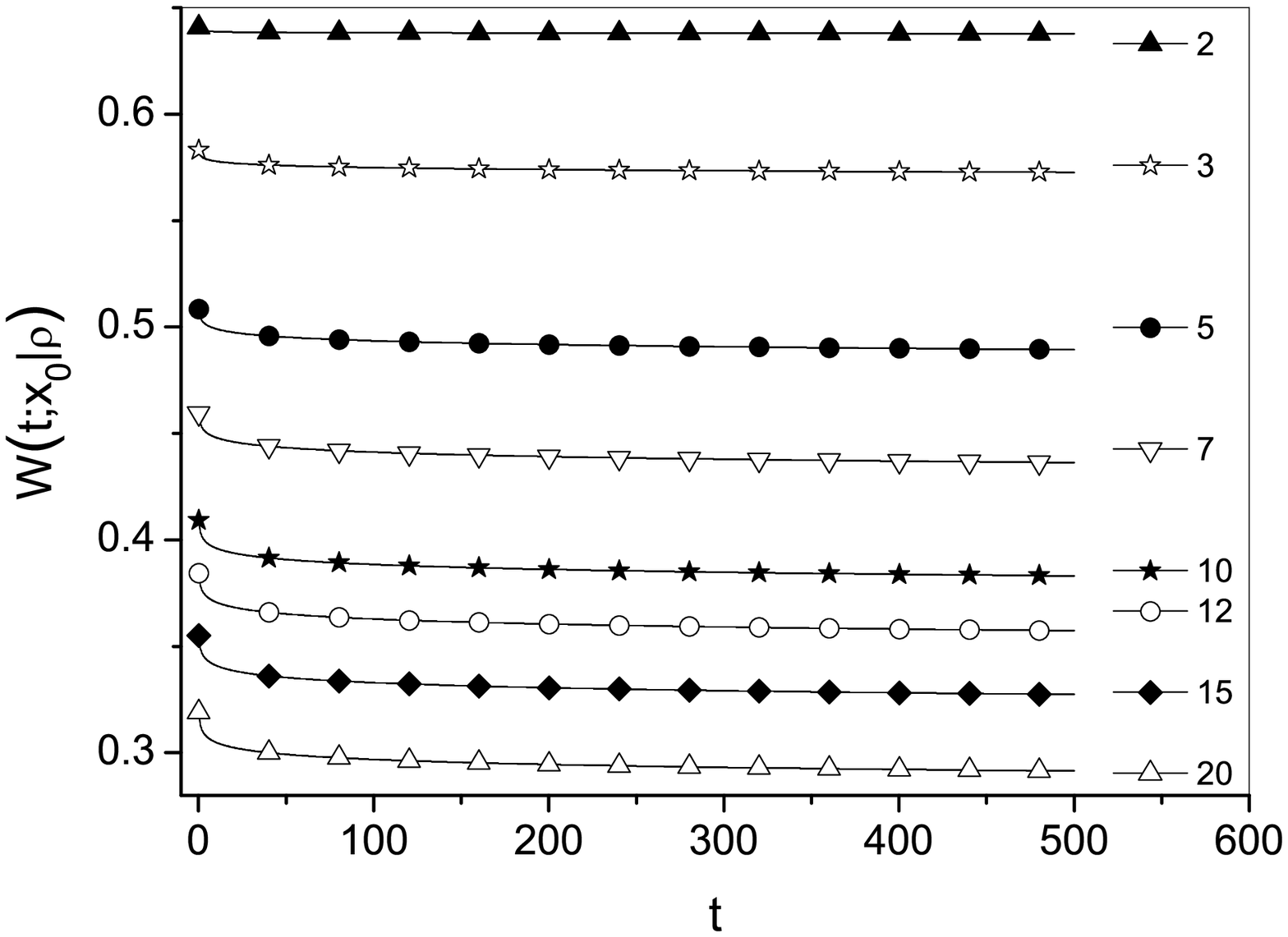}}
\caption{The plots of $W$ for different $D$ given in the legend, $\alpha=0.1$, the other parameters are the same as in Fig. \ref{Fig8}.}
\label{Fig9}
\end{figure}

Four boundary conditions are needed to solve the subdiffusion equations in regions $A$ and $B$. Two of them are $P_{AA}(-\infty,t;x_0)=0$ and $P_{BA}(\infty,t;x_0)=0$. The next two are assumed at the wall.

The subdiffusive probability flux $J$ is defined as
\begin{equation}\label{eq61}
J_{ij}(x,t;x_0|\rho)=-D\frac{\partial^{1-\alpha}}{\partial t^{1-\alpha}}\frac{\partial P_{ij}(x,t;x_0|\rho)}{\partial x},
\end{equation}
$i,j\in\{A,B\}$, the Laplace transform of the flux reads
\begin{equation}\label{eq62}
\hat{J}_{ij}(x,s;x_0|\rho)=-Ds^{1-\alpha}\frac{\partial \hat{P}_{ij}(x,s;x_0|\rho)}{\partial x}.
\end{equation}
Combining the values of the functions $\hat{P}_{AA}$, $\hat{P}_{BA}$, $\hat{J}_{AA}$, and $\hat{J}_{BA}$ calculated for $x=x_N$, 
we obtain the boundary conditions at the wall given in terms of the Laplace transform
\begin{eqnarray}\label{eq63}
\sigma_A\rho_A\hat{P}_{AA}(x_N^-,s;x_0|\rho)\\
=\left(\sigma_B+\sqrt{\frac{s^\alpha}{D}}\right)\hat{P}_{BA}(x_N^+,s;x_0|\rho),\nonumber
\end{eqnarray}
\begin{eqnarray}\label{eq64}
\rho_A\sqrt{\frac{s^\alpha}{D}}\hat{J}_{AA}(x_N^-,s;x_0|\rho)\\
=\left(\sigma_B(1-\rho_A\rho_B)+\sqrt{\frac{s^\alpha}{D}}\right)\hat{J}_{BA}(x_N^+,s;x_0|\rho),\nonumber
\end{eqnarray}
Using Eq. (\ref{eq3}) we get the boundary conditions in the time domain
\begin{eqnarray}\label{eq65}
\sigma_A\rho_A P_{AA}(x_N^-,t;x_0|\rho)=\sigma_B P_{BA}(x_N^+,t;x_0|\rho)\\
+\frac{1}{\sqrt{D}}\frac{\partial^{\alpha/2} P_{BA}(x_N^+,t;x_0|\rho)}{\partial t^{\alpha/2}},\nonumber
\end{eqnarray}
\begin{eqnarray}\label{eq66}
\rho_A \frac{\partial^{\alpha/2} J_{AA}(x_N^-,t;x_0|\rho)}{\partial t^{\alpha/2}}\\
=\sigma_B(1-\rho_A\rho_B)J_{BA}(x_N^+,t;x_0|\rho)\nonumber\\ 
+\frac{1}{\sqrt{D}} \frac{\partial^{\alpha/2} J_{BA}(x_N^+,t;x_0|\rho)}{\partial t^{\alpha/2}}.\nonumber
\end{eqnarray}

\subsection{Probability that a diffusing particle still exists in the system\label{sec4c}}

The probability $W_{AA}$ that the particle still exists in the system and is in region $A$ at time $t$ is expressed by the formula
$W_{AA}(t,x_0|\rho)=\int_{-\infty}^{x_N} P_{AA}(x,t;x_0|\rho)dx$, similar probability $W_{BA}$ for the region $B$ is
$W_{BA}(t,x_0|\rho)=\int_{x_N}^\infty P_{BA}(x,t;x_0|\rho)dx$. After calculations we get the probabilities in terms of the Laplace transform
\begin{eqnarray}\label{eq67}
\hat{W}_{AA}(s;x_0|\rho)=\frac{1}{s}\\
-\frac{1}{2s}\big[1-\Xi_{A0}(s)-\Xi_{AA}(s|\rho)\big]\;{\rm e}^{-\sqrt{\frac{s^\alpha}{D}}(x_N-x_0)},\nonumber
\end{eqnarray}
\begin{eqnarray}\label{eq68}
\hat{W}_{BA}(s;x_0|\rho)=\frac{1}{2s}\;\Xi_{BA}(s|\rho)\;{\rm e}^{-\sqrt{\frac{s^\alpha}{D}}(x_N-x_0)}.
\end{eqnarray}
The probability $W$ that the particle still exists in the system reads 
\begin{equation}\label{eq69}
W(t;x_0|\rho)=W_{AA}(t;x_0|\rho)+W_{BA}(t;x_0|\rho).
\end{equation} 
From Eqs. (\ref{eq46})--(\ref{eq48}) and (\ref{eq67})--(\ref{eq69}) we get
\begin{eqnarray}\label{eq70}
\hat{W}_{AA}(s;x_0|\rho)=\frac{1}{s}\left(1-{\rm e}^{-\sqrt{\frac{s^\alpha}{D}}(x_N-x_0)}\right)\\
+\frac{\sqrt{D}}{s}\sum_{n=0}^\infty g_n s^{\alpha(n+1)/2}{\rm e}^{-\sqrt{\frac{s^\alpha}{D}}(x_N-x_0)},\nonumber
\end{eqnarray}
\begin{eqnarray}\label{eq71}
\hat{W}_{BA}(s;x_0|\rho)=\frac{\sqrt{D}}{s}\sum_{n=0}^\infty h_n s^{\alpha(n+1)/2}\\
\times\;{\rm e}^{-\sqrt{\frac{s^\alpha}{D}}(x_N-x_0)}.\nonumber
\end{eqnarray}
From the above equations, in the time domain we have
\begin{eqnarray}\label{eq72}
W(t;x_0|\rho)=1-f_{-1,\alpha/2}\left(t;\frac{x_N-x_0}{\sqrt{D}}\right)\\
+\sqrt{D}\sum_{n=0}^\infty (g_n+h_n) f_{(n+1)\alpha/2-1,\alpha/2}\left(t;\frac{x_N-x_0}{\sqrt{D}}\right).\nonumber
\end{eqnarray}
Due to Eq. (\ref{eq45}), we obtain in the long time limit 
\begin{eqnarray}\label{eq73}
W(t;x_0|\rho)=\frac{1}{\Gamma(1-\alpha/2)\sqrt{D}t^{\alpha/2}}\Bigg[x-x_0\\
+\frac{1}{1-\rho_A\rho_B}\left(\frac{1}{\sigma_A}+\frac{\rho_A}{\sigma_B}\right)\Bigg].\nonumber
\end{eqnarray}
The plots of the function $W$ Eq. (\ref{eq72}) are presented in Figs. \ref{Fig3}--\ref{Fig9}, the parameters are given in arbitrarily chosen units. For all cases there is $x_N=0$ and $x_0=-1$. Fig. \ref{Fig3} is made for different $\alpha$ when other parameters are constant. Figs. \ref{Fig4}--\ref{Fig9} are made for two values of $\alpha$ differing significantly from each other, $\alpha=0.9$ and $\alpha=0.1$, the values of other parameters are given in the legend and figure captions. The dispersion of the values of $W$ for long times when only one parameter changes shows how strong is the influence of the parameter on $W$. 

\section{Final remarks\label{sec5}}

We have presented a model of subdiffusion in a system containing a thin partially permeable wall that can, with some probability, absorb particles diffusing through it. The probability of particle absorption depends on the number of particle passing through the wall. We do not assume where specifically the absorption of particles takes place, on the surface of the wall or inside it, but absorption can occur when a particle jumps through the wall. Although subdiffusion is qualitatively different from normal diffusion, by substituting $\alpha=1$ in the equations derived in this paper for subdiffusion we get equations for normal diffusion.

We have shown that modelling subdiffusion of a particle in a system with partially permeable partially absorbing wall can be reduced to solve the subdiffusion equation Eq. (\ref{eq1}) with boundary conditions at the wall Eqs. (\ref{eq65}) and (\ref{eq66}). The absorption properties of the wall are involved in boundary conditions at the wall. The boundary conditions are useful in solving the subdiffusion equation in a system with more than one PPAW. We recommend solving subdiffusion equations by means of the Laplace transform method, and calculating the inverse Laplace transform of the obtained solutions using the procedure described in Sec. \ref{sec4a}. The boundary conditions have the following properties:
\begin{itemize}
	\item When the particles diffuse independently of each other and do not change the properties of the wall, due to the equation $C(x,t)=\int C(x_0,t)P(x,t;x_0)dx_0$ where $C$ is a concentration of diffusing particles, the boundary conditions Eqs. (\ref{eq65}) and (\ref{eq66}) are also valid for $C(x,t)$. 
	\item When $\sigma_A=\sigma_B$, in a long time limit, which corresponds to the limit of the small parameter $s$, boundary condition Eq. (\ref{eq65}) shows that the probability of finding a particle near the membrane in region $A$ (in which the initial particle position is located) is smaller than in region $B$, $\hat{P}_{AA}(x_N^-,s;x_0)<\hat{P}_{BA}(x_N^+,s;x_0)$. The interpretation of this fact is as follows. After a sufficiently long time in both regions close to the wall one expects that, with hight probability, a particle has crossed the wall at least once. The return of the particle to $A$ requires one more particle jump through the wall comparing to the situation that the particle stays in $B$. During this jump the particle may be absorbed. Thus, the probability to find the particle at $x_N^-$ is smaller than the probability of finding the particle at $x_N^+$. 
	\item When $\sigma_B\neq 0$, the boundary condition Eq. (\ref{eq66}) shows that the flux is continuous at the wall only if absorption at the wall is absent, $\rho_A=\rho_B=1$. If $\sigma_B=0$, the transition of the particle from $B$ to $A$ is not possible and the flux is continuous at the wall when $\rho_A=1$.
	\item Boundary conditions strongly depend on subdiffusion parameter $\alpha$. In practice, this remark means that, e.g., boundary conditions at a thin membrane used for normal diffusion cannot be applied, without proper justification, at the same membrane placed in a subdiffusive medium.
\end{itemize}

The presented model assumes that the wall representing membrane is thin and the particle cannot diffuse inside the membrane. As we have mentioned in Sec. \ref{sec1}, if the system contains a thick membrane in which diffusion of particles is possible, a three-layer model of the system may be used; the middle layer represents the thick membrane. The boundary conditions derived in this paper may be assumed at the surfaces of the thick membrane. Such boundary conditions describe a possible absorption process of diffusing particles on the membrane surfaces.

A filtering process in which diffusing substance is absorbed and removed from further diffusion is described by the function $W$ which is defined as the probability that diffusing particle still exists in the system. The efficiency of the filtering process depends on the frequency of particle jumps through the wall and the absorption properties of the wall. The jump frequency depends on the subdiffusion parameters $\alpha$ and $D$, and the wall permeabilities $\sigma_A$ and $\sigma_B$.  
The dispersion of the values of the function $W$ at long time calculated for different values of the parameters, presented in Figs. \ref{Fig3}--\ref{Fig9}, suggests the following conclusions. 
For a relatively large parameter $\alpha=0.9$ the parameter $D$ has the greatest impact on the $W$ function, the impact of absorption probabilities $\rho_A$ and $\rho_B$ is slightly smaller, and the effect of the wall permeability coefficients seems to be small. For a small parameter $\alpha=0.1$ the scatter of the plots for different values of other parameters is noticing more than in the case of $\alpha=0.9$, but still the parameter $D$ has the greatest impact, the next being the wall absorption coefficients. The above conclusions obtained on the basis of the plots of the $W$ function should be treated rather as a suggestion. However, the hypothesis that the subdiffusion parameter $\alpha$ has the greatest effect on $W$ for long time seems to be well motivated. 

The considerations have been made assuming that the particle was initially in the region $A$, $x_0<x_N$. Due to the symmetry arguments, we can obtain the Green's functions for the case of $x_0>x_N$ doing the following changes of indexes $A\rightarrow B$, $B\rightarrow A$, and $(m-m_0,m-m_N,m_0-m_N)\rightarrow (m_0-m,m_N-m,m_N-m_0)$ in appropriate equations derived in this paper.

\section*{Appendix I}

The generating functions for Eqs. (\ref{eq10})--(\ref{eq13}) have been derived in \cite{tk1}. These functions read
\begin{eqnarray}\label{a1}
S_{P_{AA}}(m,z;m_0)=\frac{[\eta(z)]^{|m-m_0|}}{\sqrt{1-z^2}}\\
+\left[\frac{q_A-q_B\eta(z)}{1-(q_A+q_B-1)\eta(z)}\right]\frac{[\eta(z)]^{2N-m-m_0+1}}{\sqrt{1-z^2}}\;,\nonumber
\end{eqnarray}
\begin{eqnarray}\label{a2}
S_{P_{BA}}(m,z;m_0)=\frac{[\eta(z)]^{m-m_0}}{\sqrt{1-z^2}}\\ 
\times\frac{(1+\eta(z))(1-q_A)}{\left[1-(q_A+q_B-1)\eta(z)\right]}\;,\nonumber
\end{eqnarray}
\begin{eqnarray}\label{a3}
S_{P_{AB}}(m,z;m_0)=\frac{[\eta(z)]^{m_0-m}}{\sqrt{1-z^2}}\\ 
\times\frac{(1+\eta(z))(1-q_B)}{\left[1-(q_A+q_B-1)\eta(z)\right]}\;,\nonumber 
\end{eqnarray}
\begin{eqnarray}\label{a4}
S_{P_{BB}}(m,z;m_0)=\frac{[\eta(z)]^{|m_0-m|}}{\sqrt{1-z^2}}\\
+\left[\frac{q_B-q_A\eta(z)}{1-(q_A+q_B-1)\eta(z)}\right]\frac{[\eta(z)]^{m+m_0-2N-1}}{\sqrt{1-z^2}}\;.\nonumber
\end{eqnarray}
The function $S_{P^{(0,0)}_{AA}}$ can be obtained from $S_{P_{AA}}$ Eqs. (\ref{a1}) putting $q_B=1$ and $S_{P^{(0,0)}_{BB}}$ from $S_{P_{BB}}$ Eq. (\ref{a4}) putting $q_A=1$.

The function $f_{\nu,\beta}$ Eq. (\ref{eq45}) can be expressed by the H--Fox function
\begin{eqnarray*}
  f_{\nu,\beta}(t;a)= \frac{1}{\beta a^{(1+\nu)/\beta}} H^{1 0}_{1 1}\left(\left.\frac{a^{1/\beta}}{t}\right|
    \begin{array}{cc}
      1 & 1 \\
      (1+\nu)/\beta & 1/\beta
    \end{array}
  \right)
  \;,
\end{eqnarray*}
and by the Wright ${\rm W}$ function, 
\begin{displaymath}
f_{\nu,\beta}(t;a)=\frac{{\rm W}\left(-a/t^\beta;-\beta,-\nu\right)}{t^{\nu+1}}.
\end{displaymath}


\begin{thebibliography}{33}
\bibitem{comment} The formula for $S_F$ derived in \cite{montroll65} is similar to Eq. (\ref{eq22}) presented in this paper but contains the function equivalent to $S(m_0,m_0)$ in the denominator, and may be used to calculate the first passage time in a homogeneous system only. Eq. (\ref{eq22}) may by used for a system with a partially permeable wall.
\end{thebibliography}
\end{document}